\documentclass[aps,twocolumn,a4page,prd,superscriptaddress,floatfix,nofootinbib]{revtex4}
\usepackage{graphicx}
\begin{document}

\title{Probing dark energy beyond $z=2$ with CODEX}
\author{P. E. Vielzeuf}
\email[]{up110370652@alunos.fc.up.pt}
\affiliation{Centro de Astrof\'{\i}sica da Universidade do Porto, Rua das Estrelas, 4150-762 Porto, Portugal}
\affiliation{Faculdade de Ci\^encias, Universidade do Porto, Rua do Campo Alegre 687, 4169-007 Porto, Portugal}
\affiliation{Universit\'e Paul Sabatier---Toulouse III, 118 route de Narbonne 31062 Toulouse Cedex 9, France}
\author{C. J. A. P.  Martins}
\email[]{Carlos.Martins@astro.up.pt}
\affiliation{Centro de Astrof\'{\i}sica da Universidade do Porto, Rua das Estrelas, 4150-762 Porto, Portugal}

\begin{abstract}
Precision measurements of nature's fundamental couplings and a first measurement of the cosmological redshift drift are two of the key targets for future high-resolution ultra-stable spectrographs such as CODEX. Being able to do both gives CODEX a unique advantage, allowing it to probe dynamical dark energy models (by measuring the behavior of their equation of state) deep in the matter era and thereby testing classes of models that would otherwise be difficult to distinguish from the standard $\Lambda$CDM paradigm. We illustrate this point with two simple case studies.
\end{abstract}
\maketitle

\section{Introduction}

The observational evidence for the acceleration of the universe demonstrates that canonical theories of gravitation and particle physics are incomplete, if not incorrect. The next generation of astronomical facilities must therefore be able to carry out precision consistency tests of the standard cosmological model and search for definitive evidence of new physics beyond it.

CODEX \cite{Cristiani} is a spectrograph planned for the European Extremely Large Telescope (E-ELT). It should provide the first measurement of the cosmological redshift drift (known as the Sandage-Loeb test \cite{Sandage,Loeb}); a detailed feasibility study has been carried out by Liske {\it et al.} \cite{Liske}, and other aspects relevant for our work have been explored in \cite{Corasaniti,Balbi}. Another of its goals is an improved test of the stability of nature's fundamental couplings such as the fine-structure constant $\alpha$ and the proton-to-electron mass ratio $\mu$. Apart from the intrinsic importance of these measurements, they can be used (under certain assumptions) for detailed characterization of dark energy properties all the way up to redshift 4. This was suggested in \cite{Nunes} (see also \cite{Parkinson} for a related approach), and an assessment in the context of CODEX (and its predecessor ESPRESSO) can be found in \cite{Amendola}.

We illustrate how CODEX can probe dark energy beyond the regime where it is dominating the universe's dynamics---i.e., deep in the matter era. We introduce these two observational tools in Sect. II, and discuss them in the context of two representative classes of models in Sects. III-IV, highlighting their potential synergies. Our conclusions are in Sect. V.

\section{The observational tools}

In realistic dynamical dark energy scenarios the (presumed) scalar field should be coupled to the rest of the model, unless one postulates a (yet unknown) symmetry to suppress these couplings. The relevant coupling here is the one between the scalar field and electromagnetism, which we assume to be
\begin{equation}
{\cal L}_{\phi F} = - \frac{1}{4} B_F(\phi) F_{\mu\nu}F^{\mu\nu}
\end{equation}
where the gauge kinetic function $B_F(\phi)$ is linear, 
\begin{equation}
B_F(\phi) = 1- \zeta \kappa (\phi-\phi_0)\,,
\end{equation}
$\kappa^2=8\pi G$, and the coupling $\zeta$ is related to Equivalence Principle violations. Local constraints are (conservatively) $|\zeta_{local}|<10^{-3}$ \cite{Pospelov,Dvali}. Independent constraints can be obtained from the Cosmic Microwave Background \cite{Calabrese}, and are currently about one order of magnitude weaker. This form of $B_F(\phi)$ can be seen as the first term of a Taylor expansion, and given the tight low-redshift constraints on varying couplings and on Equivalence Principle violations it is a good approximation for the redshift range being considered.

The assumption here is that the dark energy and the varying $\alpha$ are due to the same dynamical field, as in the case of nonminimally coupled quintessence models. We will also assume a flat FRW universe with $\Omega_m+\Omega_\phi=1$, neglecting the radiation contribution since we are concerned with the low-redshift behavior. The evolution of $\alpha$ is given by
\begin{equation}\label{coupling}
\frac{\Delta \alpha}{\alpha} \equiv \frac{\alpha-\alpha_0}{\alpha_0} =\zeta \kappa (\phi-\phi_0) \,,
\end{equation}
and since the evolution of the scalar field can be expressed in terms of the dark energy properties $\Omega_\phi$ and $w$ as \cite{Nunes,Amendola}
\begin{equation}\label{darkside1}
w +1= \frac{(\kappa\phi')^2}{3 \Omega_\phi} \,,
\end{equation}
(where the prime denotes the derivative with respect to $N=\ln{a}$, $a$ being the scale factor) we finally obtain the evolution of $\alpha$ in this class of models
\begin{equation}\label{darkside2}
{\alpha / \alpha_0} (a) = 1-\zeta \int_a^{a_0}\sqrt{3\Omega_\phi(a)(1+w(a))}{d\ln{a}}\,.
\end{equation}
As expected the magnitude of the variation is controlled by the strength of the coupling $\zeta$.

The Sandage-Loeb test \cite{Sandage,Loeb} is a measurement of the evolution of the redshift drift of extragalactic objects, obtained by comparing quasar absorption spectra taken at different epochs. In any metric theory of gravity the redshift drift $\Delta z$ in a time interval $\Delta t$, or equivalently the spectroscopic velocity shift $\Delta v$ (which is the directly measured quantity) is
\begin{equation}
\Delta z=(1+z)\frac{\Delta v}{c}=\Delta t \left[H_0(1+z)-H(z)\right]\,.
\end{equation}
This provides a direct measurement of the expansion history of the universe, with no model-dependent assumptions beyond those of homogeneity and isotropy. A positive drift is a smoking gun for a dark energy component accelerating the universe; a deccelerating universe produces a negative drift.

The Lyman-$\alpha$ forest (and possibly other absorption lines, including metal ones) is ideal for this measurement, but it can only be done at redshifts $z>1.7$ (in what follows, we will assume measurements between $z=2$ and $z=5$). This applies to ground-based facilities; measurements at lower redshift would be highly desirable (since they would probe the dark energy dominated epoch), but they would need to be done from space, and there is currently no envisaged space-based spectrograph with the required resolution and stability. Liske {\it et al.} \cite{Liske} have studied in detail the performance of the envisaged CODEX spectrograph, finding that the uncertainty in the spectroscopic velocity shift is expected to behave as
\begin{equation}
\sigma_v=1.35\left(\frac{S/N}{2370}\right)^{-1} \left(\frac{N_{qso}}{30}\right)^{-1/2} \left(\frac{1+z_{qso}}{5}\right)^{-1.7}\,,
\end{equation}
where $S/N$ is the signal-to-noise of the spectra, and $N_{qso}$ and $z_{qso}$ and the number of the absorption systems and their respective redshifts. This assumes photon-noise-limited observations and holds for $z\le4$; beyond that the last exponent becomes $-0.9$. In our analysis we will assume $S/N=3000$, 40 systems uniformly divided into 4 bins at $z={2,3,4,5}$ and a time between observations of $\Delta t=20$ years.

\section{A consistency test}

Suppose that the above assumption regarding varying $\alpha$ does not hold: the dark energy is due to a cosmological constant (with $w_\Lambda=-1$), and the variation of $\alpha$ is due to some other field with a negligible contribution to the universe's energy density. The Bekenstein-Sandvik-Barrow-Magueijo (BSBM) model \cite{Sandvik} is precisely of this type (it has a varying $\alpha$ field with an energy density that is no larger than that of radiation). If one neglects the recent dark energy domination one can find an analytic solution for the behavior of $\alpha$
\begin{equation}\label{alphaBSBM}
\frac{\Delta\alpha}{\alpha}=4\epsilon N=-4\epsilon\ln{(1+z)}\,,
\end{equation}
where $\epsilon$ gives the magnitude of the variation. This is sufficient for our purposes since we are mainly be interested in the matter-era behavior, but regardless of the BSBM motivation we can take this as a phenomenological parametrization. A logarithmic redshift dependence is typical for a dilaton-type scalar field in the matter-dominated era. This must satisfy the atomic clock bounds at $z=0$. Now
\begin{equation}
\left(\frac{1}{\alpha}\frac{d\alpha}{dt}\right)_0=4\epsilon H_0=1.3 (h\epsilon) \times 10^{-17}s^{-1}\,,
\end{equation}
($h$ being the Hubble parameter in units of 100 km/s/Mpc) which according to  \cite{Rosenband} is constrained to be
\begin{equation}
\left(\frac{1}{\alpha}\frac{d\alpha}{dt}\right)_0=(-1.6\pm2.3)\times 10^{-17} yr^{-1}\,;
\end{equation}
one therefore finds $\epsilon_{\rm clocks}=(5\pm 8)\times 10^{-8}$. This is smaller than the value inferred from the published evidence for a time variation of $\alpha$ \cite{Murphy1,Murphy2} (but see also \cite{Dipole}); conservatively we will assume a variation of 2 parts per million at $z\sim4$, in which case $\epsilon_{\rm webb}\sim 3\times 10^{-7}$. Although the difference is not big given the approximations being made, it does indicate that for the two measurements to be compatible the scalar field must freeze abruptly close to the present time \cite{Memorie}; the fact that we have neglected the effect of the onset of dark energy domination will not remove this difference. 

Given the $\alpha$ variation of Eq. (\ref{alphaBSBM}), one can show \cite{Nunes} that (erroneously) assuming Eq. (\ref{coupling}) to hold would lead to the following reconstructed equation of state
\begin{equation}
w(N)=(\lambda^2-3)\left[3-\frac{\lambda^2}{w_0}\frac{\Omega_{m0}}{\Omega_{\phi0}}\exp{[(\lambda^2-3)N]}\right]^{-1}\,,
\end{equation}
where
\begin{equation}
\lambda=\sqrt{3\Omega_{\phi0}(1+w_0)}=4\frac{\epsilon}{\zeta}\,.
\end{equation}
The above assumptions and conservative assumptions on $\Omega_{\phi0}$ and $w_0$ imply
\begin{equation}
10^{-3}<\lambda<1\,.
\end{equation}
\begin{figure}
\includegraphics[scale=0.5]{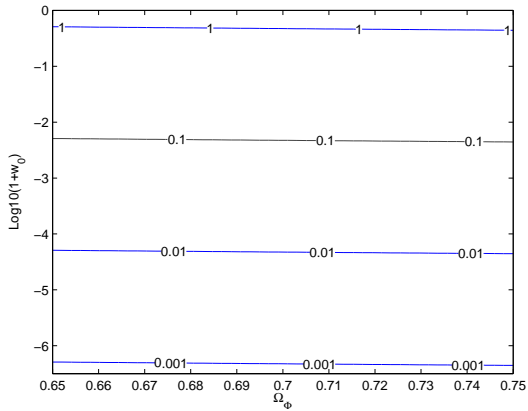}
\caption{The $\Omega_{\phi0}$-$w_0$ parameter space leading to allowed values of $\lambda$ (marked in the contour lines) compatible with a Webb-like variation of $\alpha$ and the local bounds on $\zeta$.}
\label{fig1}
\end{figure}
\begin{figure}
\includegraphics[scale=0.5]{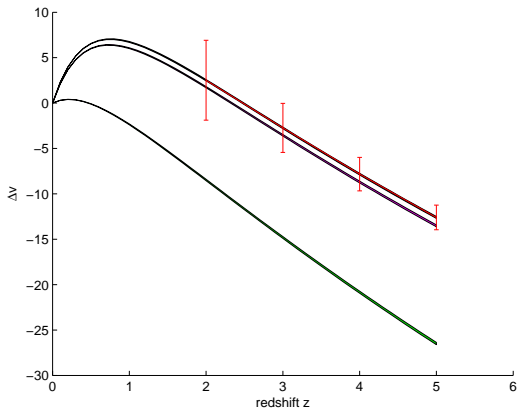}
\caption{The Sandage-Loeb signal for reconstructed BSBM models with $\lambda=1$ (bottom band) and $\lambda=0.3$ (middle), compared to the standard $\Lambda$CDM case (top band). The bands correspond the range $\Omega_{\phi0}=0.73\pm0.01$, and the vertical error bars show the uncertainty in the Sandage-Loeb signal for a CODEX dataset with an observation time of 20 years.}
\label{fig2}
\end{figure}

In fig. \ref{fig1} we show the region of the $\Omega_{\phi0}-w_0$ parameter space compatible with this range of $\lambda$, and in Fig. \ref{fig2} we plot the Sandage-Loeb signal for some representative models. Notice that for a given amount of $\alpha$ variation (that is, a value of $\epsilon$), a larger coupling $\zeta$ implies a smaller $\lambda$ (a slower-moving scalar field) and vice-versa. Large values of $\lambda$ produce a Sandage-Loeb signal that CODEX can easily distinguish from $\Lambda$CDM, which would highlight the presence of an inconsistency in the assumptions. Conversely small values of $\lambda$ yield a Sandage-Loeb signal indistinguishable from $\Lambda$CDM, but such a $\lambda$ implies a large $\zeta$ which could be checked with forthcoming (improved) local constraints. In either case, on the assumption that the current evidence for variations is correct, CODEX in combination with local experiments can support or rule out this class of models.

\section{Early dark energy}

If the dark energy is a cosmological constant its contribution to the universe's energy budget is subdominant by redshift $z\sim1$ and negligible for $z>2$. We now consider the opposite case, where the dark energy remains a significant fraction of the universe's energy density. This is realized by the 'early dark energy' models of Doran and Robbers \cite{Doran}, in which the dark energy density parameter and equation of state are
\begin{equation}
\Omega_{\phi}(a)=  \frac{\Omega_{\phi0} - \Omega_e \left(1- a^{-3 w_0}\right) }{\Omega_{\phi0} + \Omega_{m0} a^{3w_0}} + \Omega_e \left (1- a^{-3 w_0}\right)\,
\end{equation}
\begin{equation}
w(a) = -\frac{1}{3[1-\Omega_{\phi}(a)]} \frac{d\ln\Omega_{\phi}(a)}{d\ln a} + \frac{a_{eq}}{3(a + a_{eq})} 
\label{eq:edew} 
\end{equation}
$a_{eq}$ being the scale factor at matter-radiation equality; we still assume a flat Universe, so $\Omega_{m0} +\Omega_{\phi0} = 1$.

\begin{figure}
\includegraphics[scale=0.5]{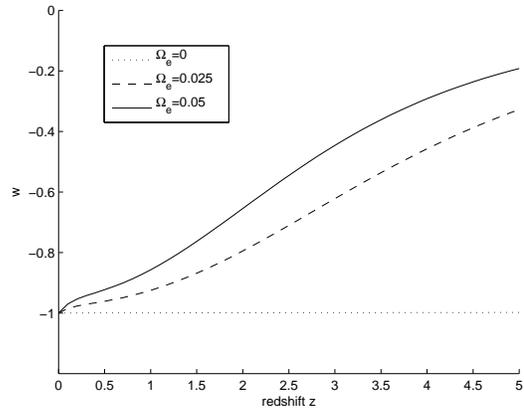}
\includegraphics[scale=0.5]{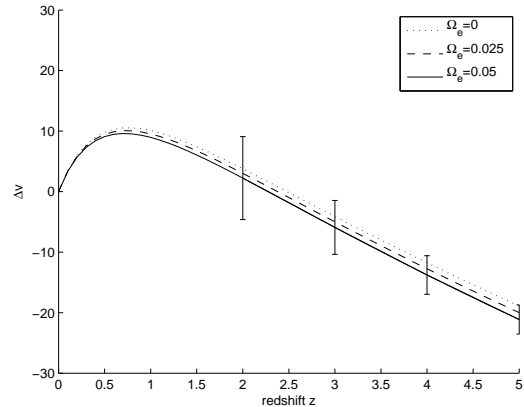}
\caption{The dark energy equation of state (top) and the Sandage-Loeb signal (bottom) for early dark energy models with $\Omega_e=0.05$ (solid) and $\Omega_e=0.025$ (dashed); the dotted line shows the standard $\Lambda$CDM. In the Sandage-Loeb plot the vertical error bars correspond to the uncertainty in the spectroscopic velocity shift for a CODEX dataset with an observation time of 20 years.}
\label{fig3}
\end{figure}
\begin{figure}
\includegraphics[scale=0.5]{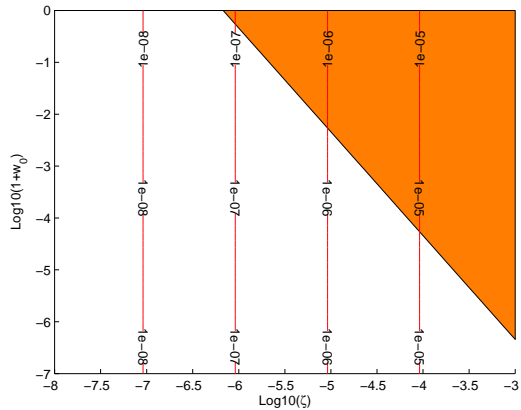}
\caption{The relative variation of the fine-structure constant, $\Delta\alpha/\alpha$, at redshift $z=4$, as a function of $\zeta$ and $w_0$ and  assuming an early dark energy model with $\Omega_e=0.05$. The shaded region is ruled out by the bound of Eq. (\protect\ref{clocks}).}
\label{fig4}
\end{figure}

The dark energy has a scaling behavior, approaching a finite constant $\Omega_e$ in the past, while its equation of state $w(a)$ tracks the dominant energy component. We assume that the early dark energy field is also coupling to electromagnetism and thus yielding a varying $\alpha$ \cite{Calabrese}, and our discussion in Sect. II, and in particular Eqs. (\ref{darkside1}-\ref{darkside2}) also apply in this case. Here our analysis is the opposite of that in the previous section: there we assumed a given amount of $\alpha$ variation; here we will assume a given amount of early dark energy, namely $\Omega_e=0.05$, consistent with current bounds allowing for possible $\alpha$ variations \cite{Calabrese}.

The local atomic clock bound is now
\begin{equation}\label{clocks}
\zeta\sqrt{3\Omega_{\phi0}(1+w_0)}<10^{-6}\,;
\end{equation}
regardless of $\Omega_e$. For $w_0$ significantly different from $-1$ this places a model-dependent constraint on $\zeta$ that is stronger than the model-independent $\zeta<10^{-3}$, but large values of $\zeta$ are possible by having $w_0$ sufficiently close to $-1$: with $\zeta\sim10^{-3}$ and $\Omega_{\phi0}\sim2/3$ the largest allowed value is $1+w_0\sim5\times10^{-7}$; this highlights the importance of atomic clock constraints, and shows that this model is almost indistinguishable from $\Lambda$CDM if one is limited to low-redshift observations.

Despite the change in the dark energy equation of state at redshift $z\sim$few, the Sandage-Loeb test is unable to distinguish this model from $\Lambda$CDM, as can be seen in Fig. \ref{fig3}. Nevertheless, this model can yield significant variations of $\alpha$, as shown in \ref{fig4}. CODEX's baseline sensitivity for $\alpha$ measurements is around the $10^{-7}$ level, and as good as $3\times 10^{-8}$ for a few ideal systems; this is enough to detect such variations for $\zeta>3\times 10^{-7}$ and use the measurements to reconstruct $w(z)$, as discussed in \cite{Amendola}.

\section{Conclusions}

We discussed two examples of CODEX's ability to probe the nature of dark energy beyond the regime where it is dynamically important and highlighted the importance of carrying out both the Sandage-Loeb test and accurate measurements of nature's fundamental couplings. All three theoretical pillars of the $\Lambda$CDM paradigm (inflation, dark matter and dark energy) rely on the presence of new, presently unknown physics. In the absence of strong indications for what this new physics is and where it can be found, it is important to search for it in multiple places, and CODEX will have a unique role to play in the $2<z<5$ redshift range.

Our analysis is simplified, but the goal is to illustrate the point at 'proof-of-concept' level. A detailed study, with precise CODEX specifications, can be done when these are finalized. Finally, we emphasize that these tests do not exist in isolation: synergies can be found with other cosmological experiments, including ESA's Euclid mission, which will probe lower redshifts. An analysis of these possibilities, in the context of a broader observational strategy, is left for future work.

\section*{Acknowledgements}

This work was done in the context of the joint Master in Astronomy of the Universities of Porto and Toulouse, supported by project AI/F-11 under the CRUP/Portugal--CUP/France cooperation agreement (F-FP02/11). We acknowledge the support of Funda\c{c}\~ao para a Ci\^encia e a Tecnologia (FCT), Portugal, through grant PTDC/FIS/111725/2009. The work of CM is funded by a Ci\^encia2007 Research Contract, supported by FCT/MCTES (Portugal) and POPH/FSE (EC).

\end{document}